\documentclass[twocolappendix]{emulateapj}
\usepackage{apjfonts,natbib,graphicx,amsmath,xspace, ulem}

\def\tc{$T_\textrm{c}$}
\def\Dt{$\Delta T$}
\def\tw{$T_\textrm{w}$}
\def\fc{$f_\textrm{c}$}
\def\lsim{\mathrel{\rlap{\lower4pt\hbox{\hskip1pt$\sim$}}
    \raise1pt\hbox{$<$}}}                
\def\gsim{\mathrel{\rlap{\lower4pt\hbox{\hskip1pt$\sim$}}
    \raise1pt\hbox{$>$}}}                
\usepackage{hyperref}

\begin{document}

\title{Using leaked power to measure intrinsic AGN power spectra of red-noise time series}

\author{S.~F.~Zhu\altaffilmark{1} and Y.~Q.~Xue\altaffilmark{1}}

\altaffiltext{1}{CAS Key Laboratory for Researches in Galaxies and Cosmology, Center for Astrophysics, Department of Astronomy, University of Science and Technology of China, Chinese Academy of Sciences, Hefei, Anhui 230026, China; zshifu@mail.ustc.edu.cn, xuey@ustc.edu.cn}

\begin{abstract}
 Fluxes emitted at different wavebands from active galactic nuclei (AGNs) fluctuate at both long and short timescales. The variation
 can typically be characterized by a broadband power spectrum, which exhibits a red-noise
 process at high frequencies. The standard method of estimating power spectral
 density (PSD) of AGN variability is easily affected by systematic biases such as red-noise leakage and
 aliasing, in particular, when the observation spans a relatively short period and is gapped.
 Focusing on the high-frequency PSD that is strongly distorted due to red-noise leakage and usually not significantly affected by aliasing,
 we develop a novel and observable normalized leakage spectrum (NLS), which describes sensitively
 the effects of leaked red-noise power on the PSD at different temporal frequencies. Using Monte Carlo simulations, we demonstrate how
 an AGN underlying PSD sensitively determines the NLS when there is severe red-noise leakage and thereby how the NLS
 can be used to effectively constrain the underlying PSD.
\end{abstract}
\keywords{galaxies: active --- methods: numerical --- methods: data analysis --- methods: statistical}

\section{Introduction}
\begin{figure}
 \centering
 \includegraphics[width=0.48\textwidth, clip]{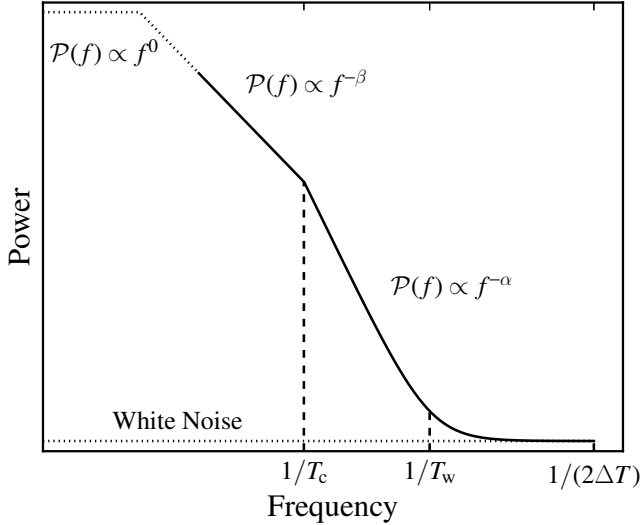}
 \caption{The broken power law that is used to model the PSD
     of AGN X-ray variability and adopted in our subsequent lightcurve simulations. $\alpha$, $\beta$, \tc\ are the
 high-frequency PSD slope, low-frequency PSD slope, and characteristic timescale, respectively; $\beta$ is fixed to 1.0 throughout the simulations.
 At 1/\tw, the power of intrinsic variability equals that of the observed white noise. \Dt\ is the time interval between consecutive samplings, and $1/(2\Delta T)$ is the highest frequency of
 the PSD, above which all power is smeared out.
 Note that since there is supposed to be a very long timescale,
 above which the variability of AGNs is no longer correlated, we plot this intrinsic white noise together with part of the low-frequency power law using dotted lines.
 Note that such a low-frequency break is physically required to make
 the variability power converge at the low-frequency end unless the low-frequency slope $\beta < 1.0$; observationally,
there is indeed one AGN whose PSD manifests this break \citep{McHardy2007}.}
\label{fig:PSD_model}
\end{figure}

Active galactic nuclei (AGNs) are powered by accretion of matter onto supermassive black holes at the centers of galaxies.
Their fluxes vary intrinsically at a wide range of timescales across a broad range of
wavebands, among which X-ray continuum emission from a postulated corona shows the most rapid aperiodic variability. Optical emission, on the
other hand, varies significantly at longer timescales. Power spectral density (PSD) is routinely used to study the distribution of variability power $\mathcal{P}(f)$\footnote{Following the notation of \cite{Vaughan2003}, we denote underlying PSD with $\mathcal{P}$ to discriminate it from periodogram that is denoted as $P$ in the subsequent text.} across temporal frequency~$f$.
    The X-ray PSD of AGNs is often modeled by a broken power law (see the sketch in Fig.~\ref{fig:PSD_model}), which usually starts with $\mathcal{P}(f) \propto 1/f^\beta$ $(\beta\sim1)$ at low frequencies, and steepens
to $\mathcal{P}(f) \propto 1/f^{\alpha}$ $(\alpha\ge 2)$, being a type of red noise, above a break frequency
\citep[e.g.,][]{Uttley2002,Markowitz2003,McHardy2004}.
The optical PSD seems to be steeper than the X-ray one at the corresponding high frequencies, with a power-law-like shape of index around or larger than 3 \citep[e.g.,][]{Mushotzky2011}.
It may also flatten to another power law with index around 2 below a break frequency \citep{Edelson2014}, and has to eventually flatten further at another low-frequency break to make power converge.

The break frequency of the X-ray PSD corresponds to a characteristic timescale, which scales linearly with mass of
black hole and possibly inversely with accretion
rate \citep{McHardy2004, McHardy2006, Kelly2011, Gonz&aacute;lez-Mart&iacute;n2012}.
This scaling relation can even extend to stellar-mass black hole accreting systems.
The amplitude of
high-frequency PSD \citep[and reference therein]{McHardy2013} or the integral of that part \citep[e.g.,][]{Zhou2010, Ponti2012} also scales inversely with
 black hole mass, which makes them potential black hole mass
 estimators. Recently, a method of measuring cosmological distance using
 amplitudes of X-ray variability of AGNs at short timescales was
 discussed \citep{La Franca2014}. In addition, the minimum timescale of
 variation is often used to constrain the size of the emitting region based on
 light crossing time arguments in the studies of X-ray \citep[e.g.,][]{Cui2004,
 Xue2005} and $\gamma$-ray \citep[e.g.,][]{Abdo2011} emission from blazars.
The minimum variation
 timescale should correspond to the end of the high-frequency power law of the PSD, if
 such most rapid flares have the same physical origin as flares of longer durations.
Multi-energy band timing studies usually reveal more physical information than monochromatic timing studies. However, verification of multi-energy band correlations and measurement of time lags between various bands still demand determination of PSD of each band \citep{Max-Moerbeck2014, Uttley2014}.

PSD, defined as the Fourier counterpart of the autocovariance function (ACF)
of a continuous process \citep{Timmer1995}, is standardly estimated by the squared-modulus of discrete Fourier
transform of an observed lightcurve, i.e., a periodogram. Because a periodogram is
not a consistent estimator of PSD \citep{Vaughan2003}, binning over frequencies
and/or segments is usually necessary \citep{van der Klis1988, Papadakis1993}. But a
binned periodogram still suffers from distortions such as red-noise leakage\footnote{For a red-noise process (e.g., $\mathcal{P}(f) \propto 1/f^{\alpha}$ with $\alpha\ge 2$), the variance (i.e., integration of $\mathcal{P}(f)$) is dominated by longterm variations. The large amount of variability power being distributed at low frequencies makes a short lightcurve display some longterm variation trend.
This longterm variability power is transferred from low frequencies to the high-frequency PSD (thus termed vividly as red-noise leakage), thereby distorting the intrinsic high-frequency PSD shape.
The severity of red-noise leakage depends both on
the shape of the underlying PSD and the length of the observation \citep[See][and reference therein]{Vaughan2003}.}
and aliasing caused by power leaking from beyond the frequency range covered by the observation.
Red-noise leakage can conceal the underlying PSD,
when the continuous
lightcurve is shorter than the characteristic timescale, which is usually the case of real AGN observations.
Such observations thus only cover the
high-frequency part of PSD, inevitably being affected by red-noise leakage.

Although several techniques have been used to alleviate this systematic distortion, the issue of red-noise leakage is far from being solved perfectly.
One such technique is called ``end matching'', which was proposed by \cite{Fougere1985}
and has still been used in many recent works \citep[e.g.,][]{Mushotzky2011,
Gonz&aacute;lez-Mart&iacute;n2012}. It removes the longterm linear trend from the lightcurve
by making the first and last data points level, with a purpose to suppress the low-frequency power leaked into the high-frequency PSD. Although
 somewhat arbitrary, this operation represents the simplest attempt to estimate
 the underlying PSD with limited observational data.

 \cite{Kelly2009} \citep[see also][]{Zu2013} proposed the use of a damped random walk (DRW) model to describe
the optical variability of AGNs. More recently, \cite{Kelly2014} developed the
continuous-time autoregressive moving average (CARMA) model to constrain
the PSD of a lightcurve. The advantage of the above two models is that they fit
the lightcurve directly, thus being free from sparse sampling.
However, some recent works \citep[e.g.,][]{Mushotzky2011, Edelson2014}
based on {\it Kepler} observations indicated that the optical high-frequency PSD of AGNs
is rather steep, deviating from a simple DRW process. The CARMA model, on the
other hand, can flexibly fit many PSDs. But the model parameters do not have
a specific physical meaning in many applications, and CARMA will always introduce
additional artificial Lorentzian features in the PSD, thus being unable to
generate the exact PSD shape as shown in Fig.~\ref{fig:PSD_model} \cite[e.g.,][]{Edelson2014}.

Another technique is called ``tapering'', i.e., using a tapered window to make
the lightcurve go to zero near its end points more gradually than a sudden drop
to zero \citep{van der Klis1988}. \cite{Max-Moerbeck2014} discussed the use of
Hanning sampling window function to reduce the effects of red-noise leakage of
unevenly-sampled time series. Their methods included Monte Carlo
simulations \citep{Timmer1995} and linear fitting of windowed periodograms. However, they
only considered PSD in the form of a simple power law, and
 thus did not discuss further the dependence of PSD
on many additional parameters such as lightcurve length and break frequency.

To resolve the issue of red-noise leakage and uncover the underlying high-frequency AGN PSD
without altering lightcurve data and without relying on
any specific physical models, we propose, in this paper, a novel and observable normalized leakage spectrum (NLS) to describe sensitively the effects of red-noise leakage on PSD and thereby constrain the
underlying AGN PSD. Note that, in this paper, we only consider a stationary process, i.e., the underlying PSD remains unchanged during observations.
The structure of this paper is as follows.
We first present the standard method to estimate PSD, the method of time-series simulation, and the effects that distort PSD in Section \ref{sec:bias}.
We then introduce the so-called NLS and examine how it is determined by the underlying PSD in Section
\ref{sec:LSall}. Subsequently, we describe our approach of constraining PSD utilizing NLS in Section
\ref{sec:method}. Finally, we discuss further NLS and its applications in Section
\ref{sec:discussion}, and summarize our results and draw our conclusions in Section \ref{sec:conclusion}.

\section{Standard method, Simulation, and Biases}
\label{sec:bias}

Continuous observations of
particular targets are usually short, lasting minutes to hours. Optical observations made
with ground-based telescopes are limited to the length of a night, or the
visibility of the target on the sky. Space telescopes may have their optical/X-ray
observations interrupted due to periodic blocking of the telescope's view
of the target by the Earth, gaps in the telemetry \citep{Vaughan2013}, or detector shutoff during
the telescope's passage over the South Atlantic Anomaly area.
As a result,
two types of observational data have to be treated differently.
A single short continuous observation corresponds to the variability observed at short timescales, i.e.,
the high-frequency PSD; while long-duration consecutive monitoring with gaps corresponds to the variability observed
at long timescales, i.e., the low-frequency PSD.
In this paper we concentrate on constraining
the underlying high-frequency PSD.
We generate artificial lightcurves using simulations with specified input PSD parameters.
The lightcurves are then sampled according to patterns similar to real observations.

\subsection{Calculating periodogram}
\label{sec:cal_per}

Periodogram is calculated as modulus-squared of the Discrete Fourier Transform of a lightcurve with an appropriate normalization.
Suppose a lightcurve $x_i$ is $T$ long and evenly sampled with interval \Dt\ at discrete
times $t_i$ ($i=1, 2,\ldots,n$), then the periodogram $P(f)$ normalized with $A_\textrm{abs}$ is:
\begin{align}
    P(f_j) &= A_\textrm{abs}|DFT(f_j)|^2 \\
 |DFT(f_j)|^2 &= \left|\sum\limits_{i=1}^{n}x_i e^{i2\pi f_jt_i}\right|^2 \nonumber \\
 &= \bigg\{\sum\limits_{i=1}^{n}x_icos(2\pi f_jt_i)\bigg\}^2 + \bigg\{\sum\limits_{i=1}^{n}x_isin(2\pi f_jt_i)\bigg\}^2
\end{align}
where $f_i=j/n\Delta T$ with $j=1,2,\ldots,n/2$. $[1/n\Delta T, 1/2\Delta T]$
is the temporal frequency range covered by the observation, where $1/2\Delta T$
is so-called Nyquist frequency.
We use absolute normalization $A_\textrm{abs}=2\Delta T/n$ \citep{Vaughan2003} so that integrating the periodogram yields $\textrm{rms}^2$
(see Section~\ref{sec:discussion} for the discussion on different normalization choices).

\subsection{Simulating artificial lightcurves}
\label{sec:sim_lc}

\begin{figure*}[!htbp]
 \centering
 \includegraphics[width=1\textwidth, clip]{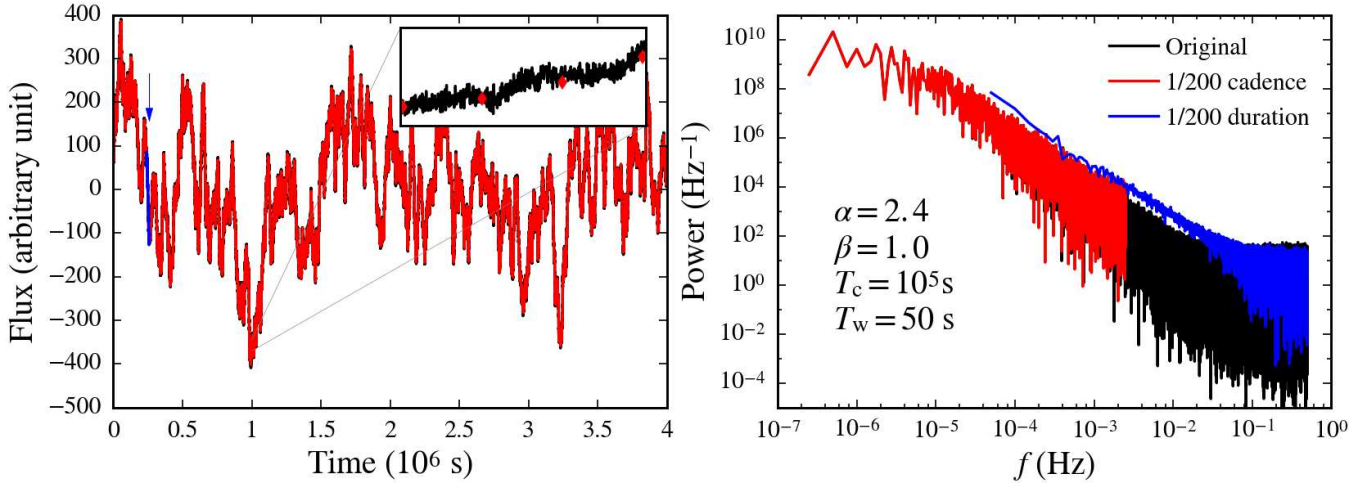}
 \caption{Left: A simulated lightcurve (in an arbitrary unit) that is $4\times10^6$s long and plotted in black, with the parameters of the corresponding PSD model listed in the right panel.
This original lightcurve is resampled with a 1/200 cadence by taking one data point out of every 200 and is plotted in red.
Note that the red lightcurve is exactly on top of the black lightcurve, so we provide a zoomed-in inset for better visualization of such a 1/200 resampling pattern.
     We also choose one random segment $2\times10^4$s long from the black lightcurve (i.e, 1/200 in length of the original black one) to represent a time-limited observation and plot it in blue
(highlighted by a downward blue arrow).
     Right: The periodograms of the three lightcurves are plotted in the same colors as the lightcurves. It is clear that the red and blue periodograms suffer from aliasing and red-noise
 leakage, respectively.}
\label{fig:two_leakage}
\end{figure*}

We use the algorithm of \cite{Timmer1995} to generate artificial lightcurves.
As plotted in Fig.~\ref{fig:PSD_model} (i.e., the solid line), a broken power law,
\begin{equation}
    \mathcal{P}(f) =
 \begin{cases}
 Af^{-\alpha} + A(1/T_\textrm{w})^{-\alpha}, & f>f_\textrm{c} \\
 Af_\textrm{c}^{-\alpha}\left({f}/{f_\textrm{c}}\right)^{-\beta}, & f<f_\textrm{c}
 \end{cases}
\label{eq:PSD_model}
\end{equation}
is used as the model PSD.
\fc\ is the break frequency that corresponds to the characteristic timescale \tc\ by $f_\textrm{c}=1/T_\textrm{c}$.
While the low-frequency logarithmic slope $\beta$ is fixed to 1.0 throughout the paper,
the high-frequency logarithmic slope $\alpha$ varies around 2.0 in the simulations.
We add a constant component to the model to represent Poisson noise.
The timescale where the power law component equals the white noise component is denoted as \tw.
$A$ is an arbitrary constant so that our simulated lightcurves are in arbitrary units.

We then multiply a complex number $(x+\textrm{i}y)/\sqrt{2}$ to the model PSD at every frequency, where $x$ and $y$ are drawn from the standard normal distribution.
The zero frequency component is set to zero, and the components along negative frequency axis are
set to the conjugate of corresponding positive frequencies (see also Appendix~B
of \cite{Vaughan2010} for another equivalent way to obtain a randomized PSD). A
subsequent inverse Fourier Transform results in an artificial lightcurve as a realization of the underlying PSD.
We note that there is a new algorithm of time series simulation developed recently by
\cite{Emmanoulopoulos2013}, which is further discussed in Section \ref{sec:discussion}.

\subsection{Red-noise leakage and aliasing}
\label{sec:two_biases}

Astronomical observations are time-limited and can only record a tiny fraction of the effectively infinitely long flux of AGNs, which
means to multiply the continuous lightcurve by a rectangular function. Such an
operation is equivalent to convoluting the PSD with the
Fourier Transform of the sampling rectangular function \citep{van der Klis1988},
which is a sinc function with strong sidelobes \citep[see Fig. 3
of][]{Max-Moerbeck2014}. Convolution transfers power from one band to
its vicinity, resulting in severe red-noise leakage when there is a significant amount of power
below the observed frequency range in, e.g., the case of a red-noise process.

Aliasing, on the other hand, happens when the observation is gapped, which means the detector is not recording the photons from the source for some reason, and the
power above the Nyquist frequency folds back. In real observations detectors integrate the flux over some smallest time
interval \Dt, which essentially smears out the high-frequency variability and prevents the power from folding back.
Moreover, the intrinsic
AGN variability drops quickly at high frequencies, where the white noise of observation
errors always dominates. Therefore, aliasing can flatten the low-frequency PSD to some degree, but is usually out of concern in the high-frequency PSD.

We show these two systematic biases of periodograms in Fig.~\ref{fig:two_leakage}.
In the left panel, we generate an artificial lightcurve (black, original) $4\times 10^6$s long with a realistic PSD model of $\alpha=2.4$, $\beta=1.0$, $T_\textrm{c}=10^5$s, and $T_\textrm{w}=50$s 
(annotated in the right panel) that is motivated by real AGN observations,
which represents a fully sampled ideal lightcurve ($\Delta T=1$s).
Then we evenly pick one out of every 200 samplings of the ideal lightcurve to obtain a discontinuously sampled
lightcurve (red, 1/200 cadence), which is 200 times undersampled. We also divide the original lightcurve into 200
segments, and choose a random one as a time-limited lightcurve (blue, 1/200 duration) so that it remains the original sampling cadence but is 200 times shorter.
The periodograms of these three lightcurves are calculated and plotted in the right panel with the colors of periodograms corresponding to the colors of lightcurves in the left panel.

As shown in the right panel of Fig.~\ref{fig:two_leakage},
the frequency ranges covered by the red and blue periodograms are narrower than the black periodogram
at the high-frequency and low-frequency ends, respectively, which is expected.
It is evident that the black periodogram well represents the underlying (model) PSD, but
the red and blue ones are clearly affected by aliasing and red-noise leakage, respectively.
Aliasing flattens the PSD of the discontinuously sampled lightcurve at the high-frequency part, i.e.,
the red periodogram is located slightly above the black periodogram around $10^{-3}$Hz (i.e., being flatter),
but the low-frequency part ($\lsim 10^{-4}$Hz) is unaffected as the red and black periodograms overlap perfectly.
The blue periodogram has an overall higher normalization and a significantly tighter dispersion compared with the black periodogram,
which indicates that, due to red-noise leakage, the periodogram of a short continuous observation does not well represent the high-frequency model PSD of a red-noise process.
At the highest frequencies ($\gsim 10^{-1}$Hz), white noise dominates for both the blue and black periodograms.

\begin{figure*}
 \centering
 \includegraphics[width=1\textwidth,clip ]{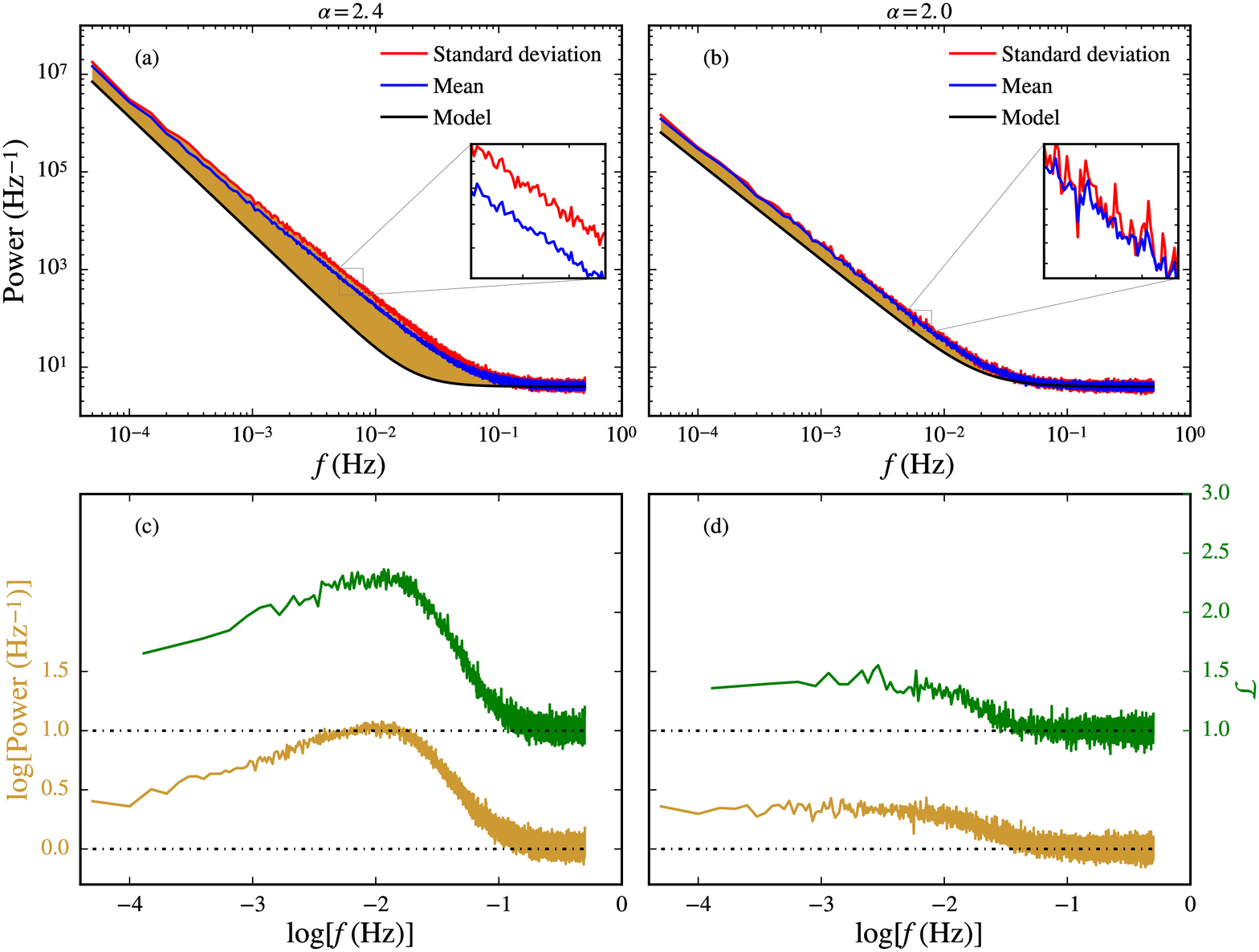}
 \caption{(a) The mean and standard deviation of the 200 periodograms (plotted as the blue and red curves, respectively) and the high-frequency model PSD (plotted as the black curve).
 The gold shaded area between the red
     and black curves shows the bias caused by red-noise leakage.
The zoomed-in inset presents more clearly the mean and standard deviation curves.
(b) Same as panel (a), but with the parameter of $\alpha=2$.
(c) The size of the gold shaded area shown in panel (a) calculated at different
     frequencies in logarithmic coordinates (plotted as the gold curve), i.e., the bias curve, with the $y=0$ dashed-dot line corresponding to the zero-bias curve (i.e., the ideal case without any red-noise leakage).
The green curve is the corresponding NLS of the 200 periodograms (see Section~\ref{sec:cal_nls} for details), with the $y=1$ dashed-dot line corresponding to the ideal NLS without any red-noise leakage.
(d) Same as (c), but corresponding to the case of panel (b).}
\label{fig:red_leakage}
\end{figure*}

\section{Normalized Leakage Spectrum}
\label{sec:LSall}

\subsection{Bias Distribution}

\label{sec:LS}

To further demonstrate the ensemble effect of red-noise leakage, in panel (a) of Fig.~\ref{fig:red_leakage}, we calculate
in linear coordinates the mean (shown as blue curve) and standard deviation (shown as red curve) of the periodograms of all the 200 segments of the original
lightcurve, and plot them together with the high-frequency part of the model PSD (shown as black curve).
According to Fig.~\ref{fig:red_leakage}a, the mean spectrum becomes flatter than the model PSD that is assumed to be in the form of $1/f^{2.4}$ for this demonstration,
due to the significant contribution of the leaked power that distributes in the form of
$1/f^2$ \citep[see][and reference therein]{Gonz&aacute;lez-Mart&iacute;n2012, Vaughan2013}.
When there is no red-noise leakage, the priodogram follows a $\chi^2(\textrm{dof}=2)$
distribution (i.e., Chi-squared distribution with two degrees of freedom) around the continuous
PSD function at every frequency \citep{van der Klis1988}: suppose that the underlying PSD is a power law with a slope $\alpha$, then the periodogram of a lightcurve is
\begin{equation}
\label{eqn:periodogram_dis}
P(f_i) = Af^{-\alpha}_{i}\frac{q_i}{2},
\end{equation}
where $q_i$ follows $\chi^2(\textrm{dof}=2)$ whose standard deviation is equal to the expectation
(i.e., 100\% dispersion). 
However, when there is red-noise leakage, the periodogram approximately becomes
\begin{equation}
\label{eqn:biased_periodogram_dis}
P(f_i) = Af^{-\alpha}_{i}\frac{q_i}{2} + Rf^{-2}_{i},
\end{equation}
where $R$ represents the amplitude of the leaked power that
is random (thus different) from a lightcurve to another.
In other words, red-noise leakage makes the distribution of power spectrum around the mean deviate from an assumed $\chi^2(\textrm{dof}=2)$ distribution,
and the resulting PSD distortion is systematic across all frequency bins; 
such a different behavior of the red-noise
distorted periodogram is
indicated by the fact that the standard deviation curve lies slightly (yet clearly) above the mean curve (see the inset of Fig.~\ref{fig:red_leakage}a).
The above deviation is caused by the fact that the leaked power is not equally distributed over different segments \citep{Uttley2002}.
An additional relevant issue, as posed in \cite{van der Klis1997}, is the shift of the bend between the power law and white noise components, i.e., $1/T_\textrm{w}$.
If \tw\ is treated as the minimum intrinsic variability timescale above the level of the observational fluctuation background, it tends to be mis-estimated from a red-noise distorted
periodogram and the resulting shift could be as large as one order of magnitude.

The gold shaded area in Fig.~\ref{fig:red_leakage}a, enclosed by the standard deviation curve and the model PSD curve, roughly shows the ``bias'' introduced by red-noise leakage.
We calculate the size of this shaded area in the logarithmic coordinates at different frequencies and plot the derived bias curve in the same color in panel (c).
The bias curve (i.e., the gold curve in Fig.~\ref{fig:red_leakage}c) rises first because of the different slopes for the model PSD and the standard deviation curve, then peaks around $1/T_\textrm{w}$, and finally falls to zero.
Apparently, a zero-bias curve corresponds to the ideal case without any red-noise leakage.
As a comparison, we repeat the above procedure and carry out another set of simulations by changing only the value of $\alpha$ to 2.0, i.e.,
    keeping all the other parameters of the model PSD the same as those adopted in Figs.~\ref{fig:red_leakage}a and \ref{fig:red_leakage}c.
We plot the corresponding results in Figs.~\ref{fig:red_leakage}b and \ref{fig:red_leakage}d.
It appears that (1) the standard deviation curve and the mean curve overlap largely although the former being somewhat higher than the latter (see the inset of Fig.~\ref{fig:red_leakage}b), and (2) the bias curve is basically flat and falls to zero without any peak.
These results are expected given that the model and biased mean periodograms have the same slope (i.e., $\alpha=2.0$).

    If we want to uncover the underlying PSD from the red-noise distorted periodogram, a seemingly obvious way would be to subtract the bias from the observed distorted periodogram.
Such an approach will, unfortunately, not work in practice because we cannot determine
    the bias without knowing the underlying PSD.
    However, what we do know is that a power spectrum unaffected by red-noise leakage should follow a $\chi^2(\textrm{dof}=2)$ distribution, whose mean is identical
    to its standard deviation; therefore,
the model PSD in this case stands for not only the expectation of a periodogram that is a realization of the model PSD,
    but also the standard deviation curve.
Due to red-noise leakage, the power of a periodogram calculated at different frequencies are no longer independent \citep{Done1992}, which results in the much tighter dispersion
of periodogram as shown in the blue spectrum in the right panel of Fig.~\ref{fig:two_leakage}, in stark contrast to the black spectrum.
On the other hand, the standard deviation curve lies above
the mean curve, which is shown in Figs.~\ref{fig:red_leakage}a and \ref{fig:red_leakage}b and indicates an additional dispersion between different segments due to red-noise leakage.
The above arguments reveal that, as far as the high-frequency part of the PSD is concerned,
even though we cannot distinguish directly between the leaked power of the long-timescale variability and the intrinsic power of the short-timescale variability,
we can utilize their different behaviors of periodogram dispersion to trace the bias, thus uncovering the underlying high-frequency PSD.

\subsection{Calculating normalized leakage spectrum}
\label{sec:cal_nls}

Suppose there are $M$ evenly-sampled lightcurves of equal length, and these lightcurves are either non-overlapping segments of a single continuous observation or sporadic observations.
    We then calculate the corresponding $M$ periodograms and bin over frequencies logarithmically according to the method of \cite{Papadakis1993}
    with a bin size of $N$, i.e., for each periodogram, we group every $N$ consecutive frequencies, take the mean of powers logarithmically as the power at the logarithmic mean frequency of those $N$ frequencies, and
    denote the result as $\textrm{log}[\overline{P_i}(f)]$, where $i=1, 2,\dots,M$. Note that \cite{Papadakis1993} required $N>20$ so that the distribution of the error of the binned periodogram
    is more Gaussian in order to fit the binned periodogram correctly, which is not required in our case because we do not intend to fit the binned power spectrum and actually the $\chi^2(\textrm{dof}=2)$ distribution assumption does
    not hold in the biased periodogram of real data. With the $M$ binned periodograms, we calculate the standard deviation at every frequency and denote it as $\textrm{std}\{\textrm{log}[\overline{P}(f)]\}$.
Subsequently, we define a novel observable quantity, the normalized leakage spectrum $\mathcal{L}(f)$, i.e., NLS, as
\begin{align}
    \label{eq:esti}
    \mathcal{L}(f) &=\frac{\textrm{std}\{\textrm{log}[\overline{P}(f)]\}}{\textrm{std}\{\textrm{log}[P]\}}, {\textrm{where}} \\
    \label{eq:std_nls}
    \textrm{std}\{\textrm{log}[P]\} &= \frac{\pi}{\textrm{ln}10}\sqrt{\frac{1}{6N}}.
\end{align}
$\textrm{std}\{\textrm{log}[P]\}$ is solely determined by $N$ and is an accurate estimation of the standard deviation of the binned periodogram if the red-noise leakage is totally absent, for
the case of a single power-law distributed PSD that is valid for the high-frequency PSD above the break frequency as shown in Fig.~\ref{fig:PSD_model} (cf. Eq.~20 of \cite{Papadakis1993} for $\textrm{std}\{\textrm{log}[P]\}$).
NLS is dimensionless and describes the severity of red-noise leakage across different frequencies.
There is an obvious fact that $\mathcal{L}(f)\gsim 1$, with the following indications:
when $\mathcal{L}(f)\sim 1$,
the observation-derived $\textrm{std}\{\textrm{log}[\overline{P}(f)]\}$ (with fluctuations) is consistent with the theoretical estimation $\textrm{std}\{\textrm{log}[P]\}$,
reflecting that the red-noise leakage at the frequency $f$ is not severe; and
$\mathcal{L}(f)>1$ is expected when the red-noise leakage is severe at the frequency $f$.

We calculate the NLS of the artificial lightcurves generated previously in Figs.~\ref{fig:red_leakage}a and \ref{fig:red_leakage}b, and plot them in green color in Figs.~\ref{fig:red_leakage}c and \ref{fig:red_leakage}d. In this case, we have $M=200$ that corresponds to the 200 segments of the original artificial lightcurve (see Section~\ref{sec:LS}),
and we adopt a bin size of $N=5$, resulting in that the NLS covers a narrower frequency range than the corresponding bias curve.
It is clear that the NLS follows the trend of the bias curve tightly irregardless of the value of $N$ (see Section~\ref{sec:dep_N}), indicating that the NLS does describe well the effects of red-noise leakage.
We note that the shape of the NLS depends on $N$ (see Section~\ref{sec:dep_N}) and is therefore not necessarily exactly the same as that of the bias curve,
although it happens to seem that the NLS in Figs.~\ref{fig:red_leakage}c and \ref{fig:red_leakage}d, if shifted vertically by $\approx -1.0$, will overlap with the corresponding bias curve.

\begin{figure}
 \centering
 \includegraphics[width=0.48\textwidth,clip ]{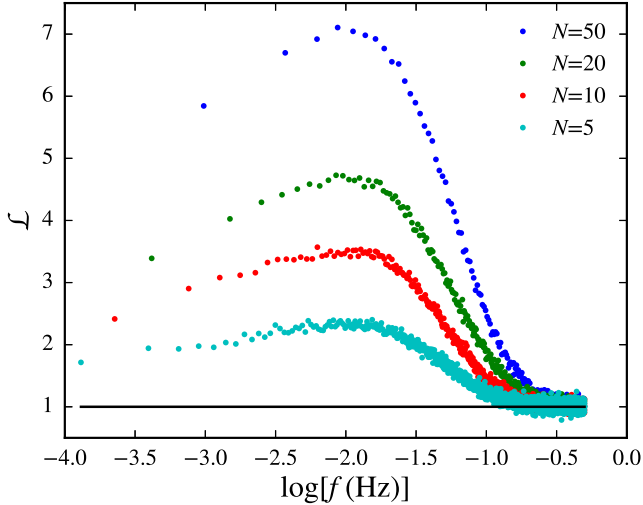}
 \caption{Dependence of NLS on the bin size of binned periodogram.
The horizontal solid line corresponds to the ideal case without any red-noise leakage.}
\label{fig:C_OnN}
\end{figure}

\subsection{Dependence on $N$}\label{sec:dep_N}

The systematic bias caused by red-noise leakage depends on the underlying PSD and the length of the observed lightcurve.
As demonstrated above, the NLS
well describes the red-noise leakage of the observations given some specific sampling pattern.
According to Eq.~\ref{eq:esti} and Eq.~\ref{eq:std_nls},
the NLS also depends on $N$, i.e., the bin size of binned periodogram.
Using the 200 segments of the original lightcurve simulated in Fig.~\ref{fig:two_leakage},
we calculate again the NLS with different $N$ values and plot the results in Fig.~\ref{fig:C_OnN}. The overall shapes of NLS with different $N$ are not the same.
The estimation of $\textrm{std}\{\textrm{log}[P]\}$, which is the denominator of the NLS, decreases with increasing $N$.
The larger $N$ is, the more the NLS deviates from the black $y=1$ horizontal solid line, making red-noise leakage seem more severe, even though its severity
is actually not determined by $N$.
The increase of $N$ not only amplifies the NLS, but also
suppresses the frequency range covered by the NLS, e.g.,
the case of $N=50$ corresponds to the narrowest frequency range of the NLS in Fig.~\ref{fig:C_OnN}.
A point worth noting is that
the rising, peaking, and declining trend is identical for the NLS with different $N$ values,
indicating that, irregardless of the $N$ values, the NLS is a reliable indicator of red-noise leakage.

\subsection{Dependence on PSD}
\label{sec:dep_psd}

\begin{figure*}
 \centering
 \includegraphics[width=1\textwidth,clip ]{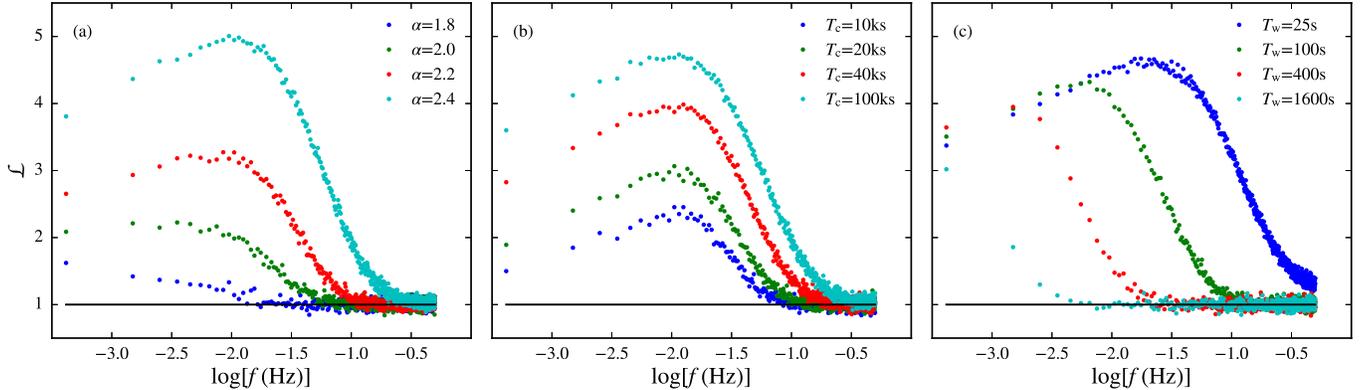}
 \caption{Dependence of NLS on the underlying PSD. The default PSD parameters are detailed in Section~\ref{sec:dep_psd}. The black $y=1$ horizontal lines
     correspond to the ideal cases without any red-noise leakage. (a) Dependence of NLS on the high-frequency slope $\alpha$ of the underlying PSD.
     (b) Dependence of NLS on the characteristic timescale \tc. (c) Dependence of NLS on \tw.
}
\label{fig:DV}
\end{figure*}

We use artificial lightcurves from simulations to examine how the NLS depends on PSD.
For each simulation, we change only one parameter out of $\alpha$, \tc, and \tw\ while
keeping the remaining PSD parameters the same as those adopted in the right panel of Fig.~\ref{fig:two_leakage} (i.e., $\alpha=2.4$, \tc$=10^5$s, \tw$=50$s, and $\beta=1.0$). The additional default parameters are
$T=4\times 10^6$s, $\Delta T=1$s, $M=200$, and $N=20$.
The results are shown in Fig.~\ref{fig:DV}.

In Fig.~\ref{fig:DV}a, only $\alpha$ is changed for each NLS. The response of the change of NLS to the change of $\alpha$ is consistent with that presented in
Figs.~\ref{fig:red_leakage}c ($\alpha=2.4$) and \ref{fig:red_leakage}d ($\alpha=2.0$), where the NLS follows closely the trend of the bias curve.
The leakage is more severe for a larger $\alpha$ value. 
Note that red-noise leakage always occurs as long
as the lightcurve is not long enough to cover all the variation power and thus shows some longterm variation trend. Even for the case of $\alpha=1.8$, the NLS still displays a sign of red-noise leakage, which is nevertheless much less severe than the cases of $\alpha\ge2$.
In general, the dependence of NLS on $\alpha$ is as follows:
when $\alpha>2$, the NLS peaks around $f=1/T_\textrm{w}$ with larger
$\alpha$ leading to higher NLS and steeper rising before the peak; when $\alpha=2$, the NLS is basically flat below $1/T_\textrm{w}$ and falls to 1 without any peak; and when $\alpha<2$, the NLS declines monotonically until
it flattens to 1.
It is also interesting to note that, below $1/T_\textrm{w}$, when $\alpha>2$ the higher-frequency part of periodogram (around $10^{-2.2}$Hz) suffers more from red-noise leakage than the lower-frequency part of periodogram (around $10^{-3.0}$Hz), i.e., the NLS slope is positive; while the trend is reversed
when $\alpha<2$ such that the lower-frequency part of periodogram suffers more from red-noise leakage than the higher-frequency part of periodogram, i.e., the NLS slope is negative.
The above results demonstrate clearly that the NLS slope (below $1/T_\textrm{w}$) has a sensitive and continuous dependence on the high-frequency slope of the underlying PSD (i.e., $\alpha$; see Fig.~\ref{fig:PSD_model}).

In Fig.~\ref{fig:DV}b, we change the characteristic timescale \tc\ and examine the dependence of NLS on it.
Apparently, \tc\ does not affect the slope of NLS below $1/T_\textrm{w}$.
Rather, \tc\ affects the height of the peak that indicates the severity of red-noise leakage.
Larger \tc\ values indicate that there is more power below the observed frequency range, thus corresponding to larger peak heights (i.e., more severe red-noise leakage).

In Fig.~\ref{fig:DV}c, we show the dependence of NLS on \tw, which represents the minimum intrinsic variability timescale that is measurable.
As the predominance of white noise starts above $1/T_\textrm{w}$, the NLS peak is always around $1/T_\textrm{w}$ for different \tw\ values.
Furthermore, because of the definition of \tw, the exact NLS peak shifts to the left of $1/T_\textrm{w}$ by about 0.1~dex (see Figs.~\ref{fig:PSD_model} and \ref{fig:red_leakage}).
Simply like \tc,
\tw\ does not affect the steepness of NLS below $1/T_\textrm{w}$.
When the S/N of the data is low, i.e., the measurable power of intrinsic variability is weak compared with the dominant white noise in the high-frequency PSD (e.g., the case of $T_\textrm{w}=1600$s),
there appears no obvious NLS peak within our interested frequency range, even though $\alpha=2.4>2$ in this case.

We conclude from Fig.~\ref{fig:DV} that the NLS depends sensitively on the underlying PSD in the following way:
$\alpha$ determines the steepness of NLS (i.e., the NLS slope below $1/T_\textrm{w}$); \tw\ determines the location of the NLS peak;
and both $\alpha$ and \tc\ affect the height of peak that indicates how severe the red-noise leakage is.
These result in the idea that we can actually use the NLS in conjunction with Monte Carlo simulations to uncover the underlying PSD of a source, which is covered only by some sporadic
observations (i.e., the typical realistic observational situation) and whose PSD is therefore subject to significant red-noise leakage.

\section{Constraining the Underlying PSD}
\label{sec:method}

In this section we demonstrate how to constrain the underlying PSD utilizing artificial lightcurves made from simulations.
In particular, we are interested in these three parameters, $\alpha$, \tc, and \tw.

\subsection{Mock observations}
\label{sec:mock_obs}

To generate an artificial lightcurve, we adopt $\alpha= 2.3$, $T_\textrm{c}=5\times10^4$s (i.e., $\textrm{log}(f_\textrm{c})=-4.70$), and $T_\textrm{w}=80$s as the input underlying high-frequency PSD, i.e., the one that we aim to uncover eventually.
We set the remaining parameters to the default values (i.e., $\beta=1.0$ and $\Delta T=1$s).
Additionally, to make our mock observations more realistic, we make the model PSD break to another white-noise level below $2\times10^{-8}$Hz (i.e., $\textrm{log}(2\times10^{-8}) =-7.70$, being
a factor of $10^3$ lower than the above high-frequency break; see Fig.~\ref{fig:PSD_model} for a schematic illustration),
given that a physically meaningful PSD has
to flatten in some way to make power converge \citep{Vaughan2003}.
Indeed, such a low-frequency break in the X-ray PSD of AGNs has been observed in, e.g., the narrow line Seyfert~1 galaxy Ark~564 \citep{McHardy2007}.
We note that this is the only simulation
in this paper where we use a double broken power-law PSD instead of a single broken power-law PSD.
From this simulation, we generate an artificial lightcurve
$1\times10^8$s long, and divide it into 5000 segments of equal duration, each lasting $2\times10^4$s.

\begin{figure}
 \centering
 \includegraphics[width=0.48\textwidth,clip]{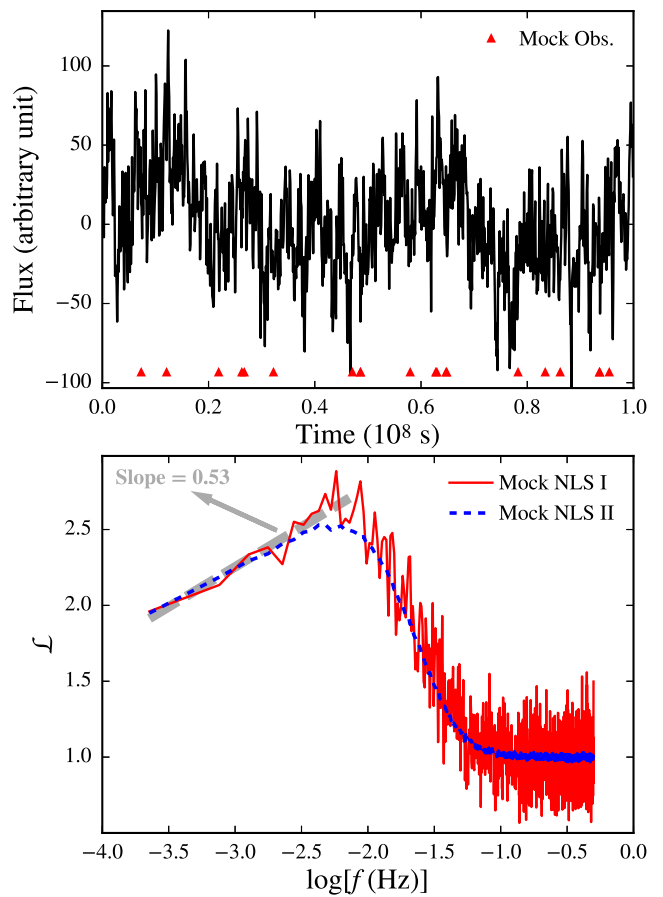}
 \caption{Top: We simulate a lightcurve $1\times10^8$s long and divide it into 5000 segments of equal duration, each lasting $2\times10^4$s. We then randomly choose 20 segments
     to mimic 20 sporadic observations, whose locations are indicated by the red triangles.
     Bottom: The NLS of the above 20 mock observations, referred to as Mock NLS I, is plotted as the red curve.
The thick grey long-dashed line indicates the best-fit linear model to the part of Mock NLS I below $1/T_\textrm{w}$, with the best-fit slope annotated.
    Another NLS, referred to as Mock NLS II, is calculated using the entire 5000 segments and plotted as the blue dashed curve.}
\label{fig:mockobs}
\end{figure}

We randomly choose 20 out of the entire 5000 segments to mimic 20 sporadic real observations.
We show the $1\times10^8$s lightcurve in the top panel of Fig.~\ref{fig:mockobs}, where the red triangles indicate the locations of the 20 random mock observations.
The NLS is calculated for these 20 short lightcurves (i.e., $M=20$) and referred to as Mock NLS I.
As Mock NLS I is sample volume limited,
we also calculate another NLS for comparison using the entire 5000 segments (i.e., $M=5000$),
in order to examine whether Mock NLS I still leads to any systematically biased result.
We refer to this later NLS as Mock NLS II,
which represents the expectation of Mock NLS I.
If we increase the number of segments included in calculating Mock NLS I, it will eventually converge to Mock NLS II.
Mock NLS I and II are shown in the bottom panel of Fig.~\ref{fig:mockobs}. Mock NLS II is much smoother than Mock NLS I because it is made using all the data available (i.e., 5000 segments) and thus is subject to much less statistical fluctuations.
The fact that Mock NLS I and II behave very similarly (i.e., they basically follow each other closely) indicates that Mock NLS I does not suffer from any appreciable systematic bias even though it makes use of only a very small portion (i.e., 20/5000=0.4\%) of the data.

\subsection{Model NLS}
\label{sec:model_VC}

We perform a large amount of additional simulations to obtain model NLS that can be
compared with the above Mock NLS I and II.
For each model PSD with a different set of parameters (i.e., $\alpha$, \tc, and \tw),
we generate a lightcurve $4\times10^7$s long,
being shorter than $1\times10^8$s adopted in the above simulation of mock observations, because the duration of a real AGN lightcurve,
which is effectively infinitely long, should not be considered as known.
If we generate the lightcurve with the same length of $1\times10^8$s, our results would only be better, at the expense of longer computational time.
Another difference is that the smallest time interval \Dt\ is 5s here, not as fine as in the mock
observations ($\Delta T=1$s), because white noise dominates the PSD at timescales smaller than \tw,
the NLS is characterless at the highest frequencies (being constant at a value of 1) while the number of data points there is large.
The adoption of a larger time resolution significantly reduces computational time.

The lightcurve is then divided into 2000 segments of equal duration, so each segment is also $2\times10^4$s long.
They are subsequently arranged into 100 groups, each of which contains 20 segments (i.e., $M=20$), the same number as the mock observations.
The above procedure eventually results in 100 NLS (i.e., each group of 20 segments resulting in a NLS),
whose mean and standard deviation at every frequency are denoted as $\overline{\mathcal{L}_{\textrm{model}}}(f)$ and $\Delta\overline{\mathcal{L}_{\textrm{model}}}(f)$.
Thus, for each given set of model PSD parameters, we can have a set of $\overline{\mathcal{L}_{\textrm{model}}}(f)$ and $\Delta\overline{\mathcal{L}_{\textrm{model}}}(f)$, which we refer to as the model NLS and its standard deviation, respectively.
By comparing the model NLS with the mock NLS (i.e., $\mathcal{L}_{\textrm{mock}}(f)$), we can examine whether the model PSD matches the underlying PSD of the mock observations.

\subsection{Assessing the goodness of `fitting'}

\begin{figure}
 \centering
 \includegraphics[width=0.48\textwidth,clip]{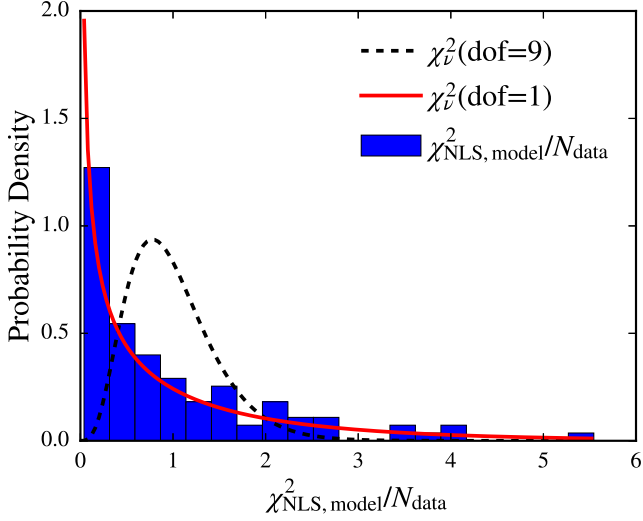}
 \caption{Normalized histogram of $\chi^2_{\textrm{NLS, model}}/N_{\textrm{data}}$ and probability density functions of two reduced Chi-squared distributions with 
9 degrees of freedom and 1 degree of freedom respectively. 
The normalized histogram of $\chi^2_{\textrm{NLS, model}}/N_{\textrm{data}}$, which is calculated using the $N_{\textrm{data}}=9$ data
points in our example (see Fig.~\ref{fig:fallex}),
does not follow the probability density function of $\chi^2_\nu(\textrm{dof}=9)$. 
Instead, due to the fact that the NLS at different frequencies
is not independent, $\chi^2_{\textrm{NLS, model}}/N_{\textrm{data}}$ follows closely the probability density function of $\chi^2_\nu(\textrm{dof}=1)$.}
\label{fig:chi_NLS_PDF}
\end{figure}

To determine how well the model NLS fits the mock NLS,
we define a special $\chi^2_{\textrm{NLS, *}}$, analogous to $\chi^2_{\textrm{dist}}$ defined in \cite{Uttley2002}:
\begin{equation}
\label{eqn:chi_squared}
\chi^2_{\textrm{NLS, *}}=
\sum\limits_{f=f_{\textrm{min}}}^{f_{\textrm{max}}}\left(\frac{\overline{\mathcal{L}_{\textrm{model}}}(f)-\mathcal{L}_{\textrm{*}}(f)}{\Delta\overline{\mathcal{L}_{\textrm{model}}}(f)}\right)^2,
\end{equation}
where the asterisk stands for mock or model, and $f_{\textrm{min}}$ and $f_{\textrm{max}}$ are the minimum and maximum frequencies of the binned periodogram of a lightcurve $2\times10^4$s long.
$\mathcal{L}_{\textrm{mock}}(f)$ is derived from the mock observations, and in our case it is either Mock NLS I or Mock NLS II, resulting in a corresponding $\chi^2_{\textrm{NLS, mock}}$ value.

The errors on NLS, being largely unknown, will in no way follow a simple Gaussian distribution.
However, using the 100 NLS available, we can determine roughly the distribution of $\chi^2_{\textrm{NLS, model}}$, by putting each of the 100 NLS into Eq.~\ref{eqn:chi_squared} and obtaining 100 different $\chi^2_{\textrm{NLS, model}}$ values.
We can then estimate, for each given set of model PSD parameters, the probability of $p(\chi^2_{\textrm{NLS, model}}>\chi^2_{\textrm{NLS, mock}})$, i.e., the total cases of $\chi^2_{\textrm{NLS, model}}$ being larger than  $\chi^2_{\textrm{NLS, mock}}$ divided by 100.
The larger $p(\chi^2_{\textrm{NLS, model}}>\chi^2_{\textrm{NLS, mock}})$ is, the better the corresponding model NLS fits the mock NLS, i.e., the better the corresponding model PSD describes the underlying PSD of the mock observations.
In this exercise, the implicit assumption is that
the unknown distribution of $\chi^2_{\textrm{NLS, mock}}$ is the same as that of $\chi^2_{\textrm{NLS, model}}$ whose PSD parameters are the same as those of
the underlying true PSD. We present the normalized distribution of $\chi^2_{\textrm{NLS, model}}/N_{\textrm{data}}$ in Fig.~\ref{fig:chi_NLS_PDF}, which shows that,  
due to the fact that the NLS at different frequencies is not independent,  
the ``effective'' degree of freedom of $\chi^2_{\textrm{NLS, model}}$ remains 1 despite of $N_{\textrm{data}}=9$ in this demonstration.
Note that we do not use bootstrapping as in \cite{Uttley2002} to determine the distribution of $\chi^2_{\textrm{NLS}}$,
because the NLS follows a non-Gaussian distribution, and the NLS at every frequency is not independent (just like the case of the red-noise distorted periodogram).
As such, bootstrapping will broaden the spread of $\chi^2_{\textrm{NLS}}$ and
result in large $p(\chi^2_{\textrm{NLS, model}}>\chi^2_{\textrm{NLS, mock}})$ values for many model PSDs. Actually, due to the same reason, $\Delta\overline{\mathcal{L}_{\textrm{model}}}$ is
larger than the actual spread of the NLS.

\subsection{Freeing one parameter}
\label{sec:free1p}

\begin{figure*}
 \centering
 \includegraphics[width=0.96\textwidth,clip]{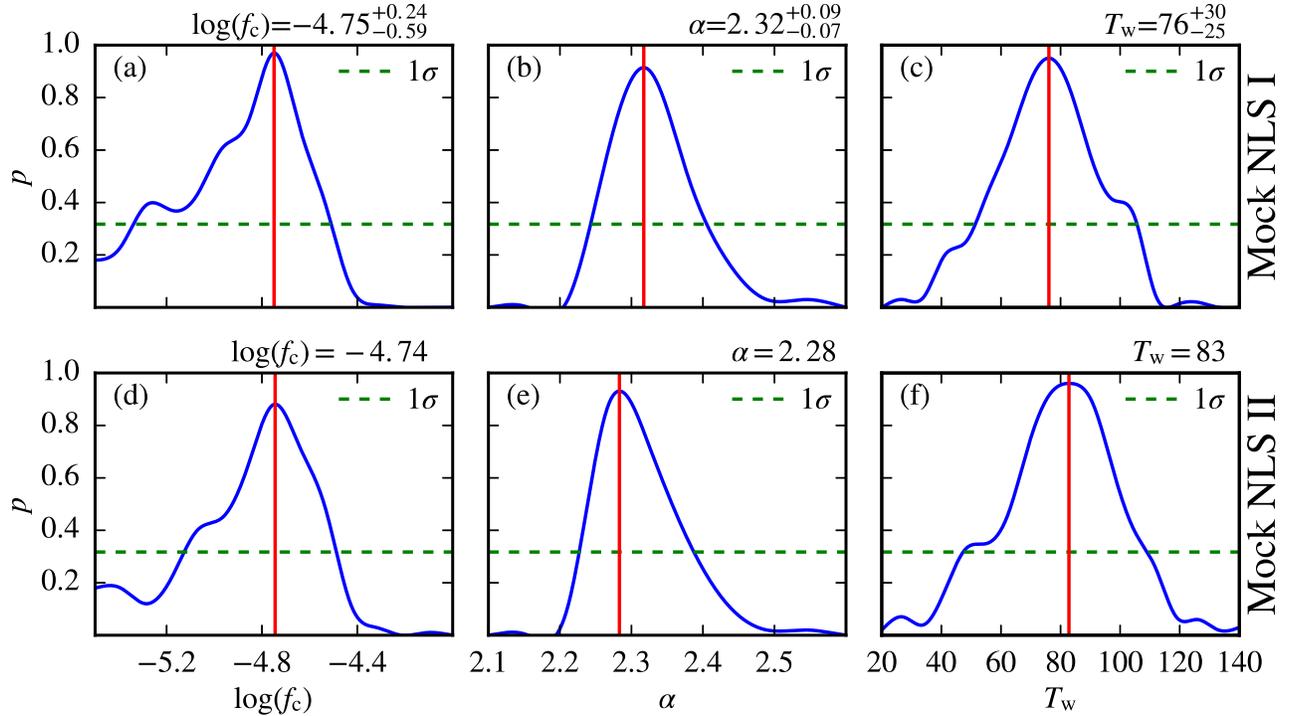}
\caption{Constraining the underlying model PSD when only one parameter is unknown (i.e., being free). The results for the case of Mock NLS I are shown in panels (a--c), and those for the case of Mock NLS II are shown in panels (d--f).
  In each panel, the $x$-axis indicates the specific unknown parameter to be constrained; the $y$-axis $p$ stands for $p(\chi^2_{\textrm{NLS, model}}>\chi^2_{\textrm{NLS, mock}})$ with the blue $p$ curve being smoothed by a cubic function;
the horizontal dashed line corresponds to $1-p=0.68$ and its intersections with the $p$ curve indicate the
$1\sigma$ range of the unknown parameter; and the best value and its $1\sigma$ range determined for the unknown parameter are denoted atop.}
\label{fig:free1p}
\end{figure*}

For the three model PSD parameters of $\alpha$, \tc, and \tw, we fix two of them to the assumed values that are used to generate mock lightcurves in Section~\ref{sec:mock_obs},
and search for the best value for the remaining ``unknown'' parameter.
The results are shown in Fig.~\ref{fig:free1p} and Table~\ref{tab:free1p}.
In real applications, these fixed parameters could be determined from low-frequency PSD modelings and/or other ways of constraints.
We do not provide the errors on the parameters in the case of Mock NLS II since it comes from a much larger set of
samples ($M=5000$) and is therefore expected to have smaller corresponding errors on the parameters than those in the case of Mock NLS I.

\begin{figure*}
 \centering
 \includegraphics[width=0.7\textwidth,clip]{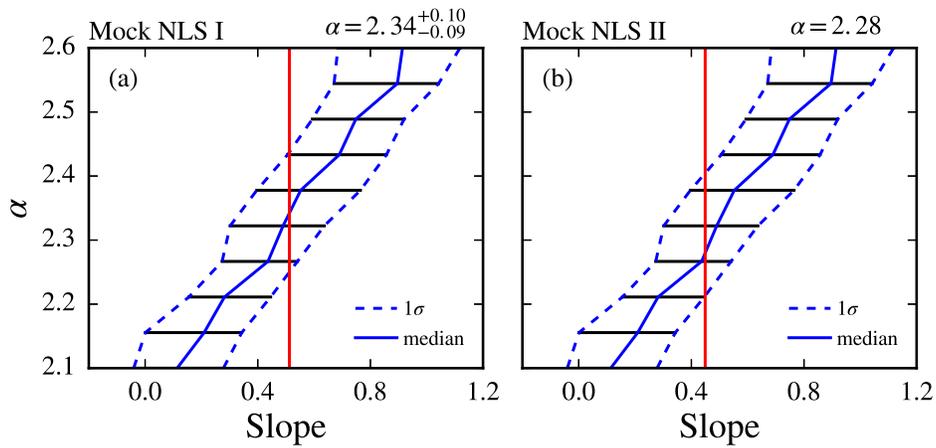}
\caption{Constraining $\alpha$ of the underlying model PSD with a different method for the cases of (a) Mock NLS I and (b) Mock NLS II, respectively.
The intersections of the vertical red lines with the
three percentile curves indicate the best $\alpha$ value and its $1\sigma$ confidence interval
(see Section~\ref{sec:free1palpha} for details).}
\label{fig:free1p2}
\end{figure*}

\begin{table}[htbp]
\caption{Results of constraining the underlying PSD when only one parameter is free}
\centering
\begin{tabular}{ccccc} \hline \hline
 Free   & Mock NLS I & Mock NLS II & Input Model \\
 Parameter & (Figure Panel) & (Figure Panel) & Parameter \\
 \hline
 $\alpha$ & $2.32_{-0.07}^{+0.09}$ (\ref{fig:free1p}b) & 2.28 (\ref{fig:free1p}e) & 2.30  \\
 $\alpha$ & $2.34_{-0.09}^{+0.10}$ (\ref{fig:free1p2}a) & 2.28 (\ref{fig:free1p2}b) & 2.30\\
 log($f_\textrm{c}$) & $-4.75_{-0.59}^{+0.24}$ (\ref{fig:free1p}a) & $-4.74$ (\ref{fig:free1p}d) & $-4.70$\\
 \tw\ & $76_{-25}^{+30}$s (\ref{fig:free1p}c) & $83$s (\ref{fig:free1p}f) & 80s \\ \hline
\end{tabular}
\label{tab:free1p}
\end{table}

\subsubsection{Freeing $\alpha$}
\label{sec:free1palpha}

We choose 10 different $\alpha$ values that are evenly spaced between 2.1 and 2.6 (i.e., $\Delta \alpha \simeq 0.044$), given that an AGN PSD has a typical value of $\alpha>2$.
In practice, without performing any simulation, one can easily determine $\alpha>2$ or $\alpha<2$ by checking whether there exists a peak in the NLS derived from observations (see Section~\ref{sec:dep_psd} and Fig.~\ref{fig:DV}a).
We show the constraints on $\alpha$ in Figs.~\ref{fig:free1p}b and \ref{fig:free1p}e for the cases of Mock NLS I and II, respectively (also see Table~\ref{tab:free1p}).
It is clear that, in both cases, the derived $\alpha=2.32_{-0.07}^{+0.09}$ and $\alpha=2.28$ are in excellent agreement with the input $\alpha=2.30$ of the model PSD.

As shown in Figs.~\ref{fig:red_leakage},
\ref{fig:DV}, and \ref{fig:mockobs}, the NLS slope below $1/T_{\rm w}$ uniquely reflects the
steepness of the underlying PSD (i.e., $\alpha$), irregardless of variations of \tc\ and \tw.
So we develop another method to constrain
$\alpha$, which was
inspired by \cite{Max-Moerbeck2014}.
From the 100 NLS derived with each $\alpha$ value, we find three percentiles, i.e., 16\%, 50\%, and 84\% of the NLS slopes, plot them in Fig.~\ref{fig:free1p2},
interpolate the same percentiles for different $\alpha$ values (i.e., the blue solid and dashed curves), and finally draw the NLS slope of Mock NLS I or II (e.g., in Fig.~\ref{fig:free1p2}a, we draw the vertical red line of slope=0.53 for Mock NLS I; this slope value is also shown in the bottom panel of Fig.~\ref{fig:mockobs}).
As shown in Figs.~\ref{fig:free1p2}a and \ref{fig:free1p2}b, the intersections of the vertical red lines with the
three percentile curves result in estimates of $\alpha$ and its $1\sigma$ confidence interval.
The derived $\alpha=2.34_{-0.09}^{+0.10}$ for Mock NLS I and $\alpha=2.28$ for Mock NLS II are again in
excellent agreement with the input $\alpha=2.30$ of the model PSD (also see Table~\ref{tab:free1p}).

Using the above two methods, we constrain well the $\alpha$ values for the cases of both Mock NLS I and Mock NLS II, with the results for Mock NLS II being better and not obviously biased.

\subsubsection{Freeing \tc}

We choose 15 different log($f_\textrm{c}$) values that are evenly spaced between $-5.5$ to $-4.0$ (i.e., $\Delta \textrm{log}(f_\textrm{c}) \simeq 0.11$).
We show the constraints on log($f_\textrm{c}$) in Figs.~\ref{fig:free1p}a and \ref{fig:free1p}d for the cases of Mock NLS I and Mock NLS II, respectively (also see Table~\ref{tab:free1p}).
It is clear that, in both cases, the derived $\textrm{log}(f_\textrm{c})=-4.75_{-0.59}^{+0.24}$ and $\textrm{log}(f_\textrm{c})=-4.74$ are in excellent agreement with the input $\textrm{log}(f_\textrm{c})=-4.70$ of the model PSD.

\subsubsection{Freeing \tw}

We choose 18 different \tw\ values that are evenly spaced between 20s and 140s (i.e., $\Delta T_\textrm{w} \simeq 7$s).
We show the constraints on \tw\ in Figs.~\ref{fig:free1p}c and \ref{fig:free1p}f for the cases of Mock NLS I and Mock NLS II, respectively (also see Table~\ref{tab:free1p}).
It is clear that, in both cases, the derived $T_\textrm{w}=76_{-25}^{+30}$s and $T_\textrm{w}=83$s are in excellent agreement with the input $T_\textrm{w}=80$s of the model PSD.

\subsection{Fixing one parameter}
\label{subsec:fix_one_parameter}

\begin{figure}
 \centering
 \includegraphics[width=0.48\textwidth,clip]{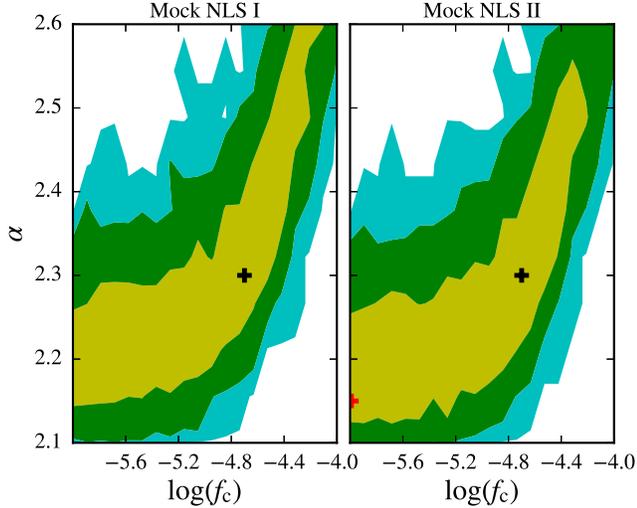}
 \caption{Contour plots of constraining simultaneously
$\alpha$ and \tc\ of the underlying model PSD
for the cases of Mock NLS I and II, when \tw\ is fixed to 80s.
The yellow, green, and cyan contours indicate $1\sigma$, $2\sigma$, and $3\sigma$ confidence intervals of $\alpha$ and log($f_{\rm c}$), whose true values in the input model PSD are marked by
     black crosses. The red cross represents an example to be displayed in Fig.~\ref{fig:fallex}
for further discussions.}
\label{fig:fixbend}
\end{figure}

In this subsection, we fix \tw\ to 80s and search for the best combination of \tc\ and $\alpha$ values.
We create a $20\times10$ grid of log($f_\textrm{c}$) and $\alpha$, with log($f_\textrm{c}$) varying from $-6.0$ to $-4.0$ ($\Delta \textrm{log}(f_\textrm{c}) \simeq 0.105$)
and $\alpha$ varying from 2.1 to 2.6 ($\Delta \alpha \simeq 0.044$), respectively.
We show the simultaneous constraints on both \tc\ and $\alpha$ using contour plots in Fig.~\ref{fig:fixbend}
for the cases of Mock NLS I and Mock NLS II.
As shown in Fig.~\ref{fig:fixbend}, for both cases, we cannot constrain well \tc\ and $\alpha$ simultaneously (i.e., the constraints are not sufficiently stringent). The shapes of the $1\sigma$, $2\sigma$, and $3\sigma$ contours are all banana-like,
which is caused by the degeneracy among various \tc-$\alpha$ combinations, i.e.,
a PSD with a larger \tc\ value and a smaller $\alpha$ value and
a PSD with a smaller \tc\ value and a larger $\alpha$ value
can have very similar and even indistinguishable red-noise leakage effects.
However, when we use the method of the NLS slope
to constrain $\alpha$ as in Section~\ref{sec:free1palpha} and Fig.~\ref{fig:free1p2},
we still obtain a well-constrained value of $\alpha=2.36_{-0.08}^{+0.10}$ for the case of Mock NLS I.
Therefore, we are able to uncover the $\alpha$ value of the underlying PSD even though \tc\ is not fixed or close to the input model value.
We do not intend to constrain the three parameters (i.e., $\alpha$, \tc, and \tw) simultaneously.

\section{Discussion and Perspective}
\label{sec:discussion}

\begin{figure}
 \centering
 \includegraphics[width=0.48\textwidth,clip]{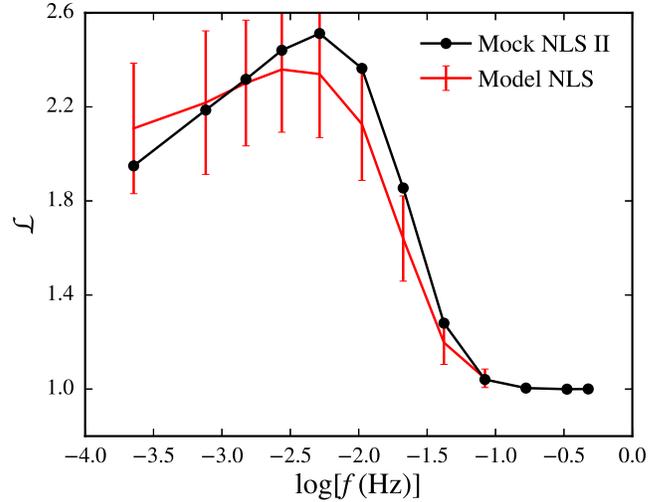}
 \caption{Comparison between the binned Mock NLS II and a model NLS with log($f_\textrm{c})=-6$ and $\alpha=2.15$. Although the PSD parameters of this model NLS deviate significantly from those of the input model PSD (see the red and black crosses in the right panel of Fig.~\ref{fig:fixbend}),
it appears that this model NLS is statistically consistent with Mock NLS II, i.e., our approach cannot rule out this particular model NLS.}
\label{fig:fallex}
\end{figure}

In this paper, our simulated lightcurves have arbitrary units,
which are generated following the assumed corresponding underlying PSDs (see Eq.~\ref{eq:PSD_model}).
We then use an absolute normalization $A_{\textrm{abs}}=2\Delta T/n$ for the calculation of a periodogram such that the integration of the periodogram equals rms$^2$ (see Section~\ref{sec:cal_per}).
Real AGN lightcurves have
some interesting properties, e.g., the log-normal distribution of fluxes
and the rms-flux relation in both the X-ray \citep{Uttley2005a} and optical \citep{Edelson2013} bands,
which calls for the use of a physically well-motivated rms-normalization $A_{\textrm{rms}}=2\Delta T/\bar{x}^2n$ in calculating the NLS based on real observational data.
In the case of using $A_{\textrm{rms}}$, the NLS will finally flatten at some value larger than 1 instead of flattening at 1, which, however, does not indicate red-noise leakage affecting the highest-frequency PSD, and is actually
due to that the Poisson noise level and thus the \tw\ value vary for lightcurve segments with different mean fluxes.
An additional normalization $A_{\textrm{Leahy}}=2\Delta T/\bar{x}n$ \citep{Leahy1983} can be considered as well because
in this case the Poisson noise level is constant and the NLS falls to 1 at high frequencies.
%
Recently, \cite{Emmanoulopoulos2013} developed a new algorithm of time series simulation,
which combines the properties of PSD and probability distribution
function (PDF) in generating a single artificial lightcurve.
Their lightcurves can produce the rms-flux relation if a log-normal PDF is adopted.
They also
re-randomized the fluxes to mimic Poisson fluctuations during observations. All these features
make the \cite{Emmanoulopoulos2013} algorithm a better way to generate artificial lightcurves. However, this algorithm is obviously much more computationally intensive.
Therefore, we do not adopt it because we are only interested in uncovering the underlying high-frequency PSD using the basic NLS properties, and our simpler method of simulating lightcurves suffices for our purpose.
Another way to reconcile the seeming contradictories at choosing different normalizations is that one could take
the logarithm of fluxes \citep{Uttley2005a, Kelly2009} instead of using count rates or fluxes \citep{Uttley2005b} that are usually adopted.
The logarithms of AGN fluxes typically distribute in a Gaussian form, thus one does not have to resort to the \cite{Emmanoulopoulos2013} method.
In the appendix, we carry out an additional set of simulations
to investigate the effects of log-normal flux distribution and Poisson-statistics sampled lightcurves (i.e., features of real AGN data) on the performance of NLS, following the method outlined in \cite{Uttley2005a}.
We find that, with an appropriate choice of normalization, the NLS can also be used to effectively 
constrain the high-frequency PSD of lightcurves with realistic features, which further
demonstrates that the NLS is an effective and consistent tool to study AGN PSD.

We use a broken power law with a fixed value of $\beta=1.0$ as our base PSD model;
nevertheless, our NLS approach can be applied to any arbitrary PSD model, e.g., the one with a smoothly bending power
law or multiple Lorentzian components \citep{McHardy2004, McHardy2007}.
Furthermore, our approach is applicable to both X-ray and optical timing observations of AGNs.
If $\beta$ is set free,
because it can affect the severity of red-noise leakage as much as \tc, our approach cannot constrain $\beta$ and \tc\ at the same time.
In other words, the NLS approach by itself cannot constrain more than one parameters of the low-frequency PSD simultaneously.
An approach of constraining a broadband PSD is described in \cite{Uttley2002}, who utilized lightcurves observed with different sampling
patterns across different timescales and from different telescopes.
However, both the \cite{Uttley2002} approach and its recently revised version \citep{Marshall2015} do not make enough efforts to alleviate the issue of red-noise leakage at the high-frequency PSD, which is the focus of our work.

We use an example in Fig.~\ref{fig:fallex} to show that simply fitting Mock NLS II with the model NLS cannot well constrain two parameters (i.e., $\alpha$ and log(\fc)) of the input model PSD simultaneously.
Although the model of $\alpha=2.15$ and log(\fc)$=-6$ deviates significantly from the input model of $\alpha=2.30$ and log(\fc)$=-4.70$ as shown in Fig.~\ref{fig:fixbend},
and one can even somehow visually distinguish the two NLS resulted from these two models in Fig.~\ref{fig:fallex},
the model of $\alpha=2.15$ and log(\fc)$=-6$ cannot be ruled out statistically,
due to that larger $\alpha$ and \tc\ values both lead to more severe red-noise leakage (i.e.,
there is some degeneracy among various combinations of $\alpha$ and \tc).
Because the NLS values at different frequencies are not independent, the error bars on the model NLS shown
in Fig.~\ref{fig:fallex} are overestimated, being inappropriately large.
However, if we can fix the value of either $\alpha$ or log(\fc), the situation will be reduced to the case of constraining one parameter at a time
where stringent constraints can be obtained (see Section~\ref{sec:free1p}).
For example, we can first set \tc\ free and nail down $\alpha$ using the approach of the NLS slope as presented in Section~\ref{subsec:fix_one_parameter};
or, we can first constrain \tc\ (or more generally speaking, the low-frequency PSD) using longterm monitoring observations.

The DRW (or the first-order continuous-time autoregressive process, or the Ornstein-Uhlenbeck
process) model \citep{Kelly2009} is widely used to model quasar
optical variabilities. Techniques based on the DRW model have been developed for
reverberation mapping \citep{Zu2011}, AGN selection \citep[e.g.,][]{MacLeod2011}, and many other applications. The
PSD of the DRW model is a Lorentzian centered at zero, being a power law with
$\alpha=2$ at high frequencies. However, some high-quality data have shown PSD results deviating from the expectation of a
simple DRW model \citep[e.g.,][]{Mushotzky2011,Kasliwal2015}. In
Fig.~\ref{fig:DV}a, we show that the NLS slope provides a diagnosis on whether the
high-frequency PSD is in the form of $\mathcal{P}(f)\propto 1/f^2$.
So we can examine whether the simple DRW model can describe the observations adequately by simply checking whether the
NLS slope is flat.

The issue of red-noise leakage cannot be solved by improving S/N of observations that can be easily achievable using some future large facilities.
Higher S/N observations potentially allow for revealing the even higher-frequency PSD
that dictates intrinsic AGN variabilities at shorter timescales, but as red-noise leakage becomes more severe at higher frequencies when $\alpha>2$, the PSD results will still be biased if not more severely.
In fact, red-noise leakage is more of an issue of sampling, which is typically limited by the length of a night or the period of an orbit.
Due to this reason,
even the proposed next-generation X-ray timing observatories such as the Large Observatory for
X-ray Timing (LOFT; \cite{De Rosa2015}) and ground-based optical survey telescopes such as the Large Synoptic Survey Telescope (LSST), both characterized by unprecedented S/N, still have to confront the issue of
red-noise leakage in the high-frequency PSD. In particular, the duration of continuous exposure of the sensitive Large Area Detector onboard LOFT is limited to $\sim3$~ks,
because there is usually a $\sim2$~ks gap due to the occultation of the Earth in a $\sim5$~ks orbital period.
Fortunately,
our approach can mimic the real sampling patterns of observations
and uncover the underlying high-frequency AGN PSD without demanding alternation on observational data, thereby offering great help in casting light on the physical origins of AGN variabilities.

\section{Summary \& Conclusions}
\label{sec:conclusion}

In this paper,
we focus on constraining the underlying high-frequency AGN PSD that is characterized by the three parameters of $\alpha$, \tc, and \tw\ (see Fig.~\ref{fig:PSD_model}), and is typically strongly distorted due to red-noise leakage (see the right panel of Fig.~\ref{fig:two_leakage}).
We develop a novel and observable NLS (see Section~\ref{sec:cal_nls} and Fig.~\ref{fig:red_leakage}) that can describe sensitively the effects of leaked red-noise power on the PSD at different temporal frequencies.
We utilize Monte Carlo simulations to show detailedly how an AGN underlying PSD determines the NLS when there is severe red-noise leakage (see Section~\ref{sec:dep_psd} and Fig.~\ref{fig:DV}) and thereby how the NLS can be used to uncover the underlying PSD (see Section~\ref{sec:method}).
In particular, we find that, when constraining only one of the above three parameters at a time, we can obtain stringent constraints on that parameter of interest.
When constraining $\alpha$ and \tc\ simultaneously, we obtain loose constraints on both parameters due to the degeneracy among various combinations of $\alpha$ and \tc. However,
using the method of the NLS slope still enables us to obtain stringent constraints on $\alpha$, irregardless of various \tc\ values.
Furthermore, in real applications of our NLS approach, we can often reduce a complicated case of
constraining two parameters simultaneously to a simple case of constraining one parameter at a time, by setting some parameter to some fiducial value through delving into all information provided by all observations available.
As such, our NLS approach can mimic the real sampling patterns of observations and
uncover the underlying high-frequency AGN PSD, without demanding interpolation, end matching, windowing, or any other alternation on observational data.
Additionally, our approach is applicable to both X-ray and optical timing observations of AGNs.
Therefore, our NLS approach appears to be a useful and effective tool to help probe
the phenomenon and eventually the physical origins of AGN variabilities
in the course of the incoming era of time domain astronomy.

\vspace{0.5cm}

We thank our astrophysical referee Dr. Phil Uttley
and an anonymous statistical referee for the helpful feedback that improved this work.
SFZ and YQX acknowledge support of
the National Thousand Young Talents program (KJ2030220004),
the 973 Program (2015CB857004),
USTC startup funding (ZC9850290195),
the National Natural Science Foundation of China (NSFC-11473026, 11421303),
the Strategic Priority Research Program
``The Emergence of Cosmological Structures''
of the Chinese Academy of Sciences (XDB09000000),
and the Fundamental Research Funds
for the Central Universities (WK3440000001).

\appendix

\section{Effects of log-normal flux distribution and Poisson-statistics sampled lightcurves on NLS}\label{sec:appendix}

\begin{figure}[!htbp]
\figurenum{A1}
\label{fig:Uttley05_lc}
 \centering
 \includegraphics[width=0.48\textwidth, clip]{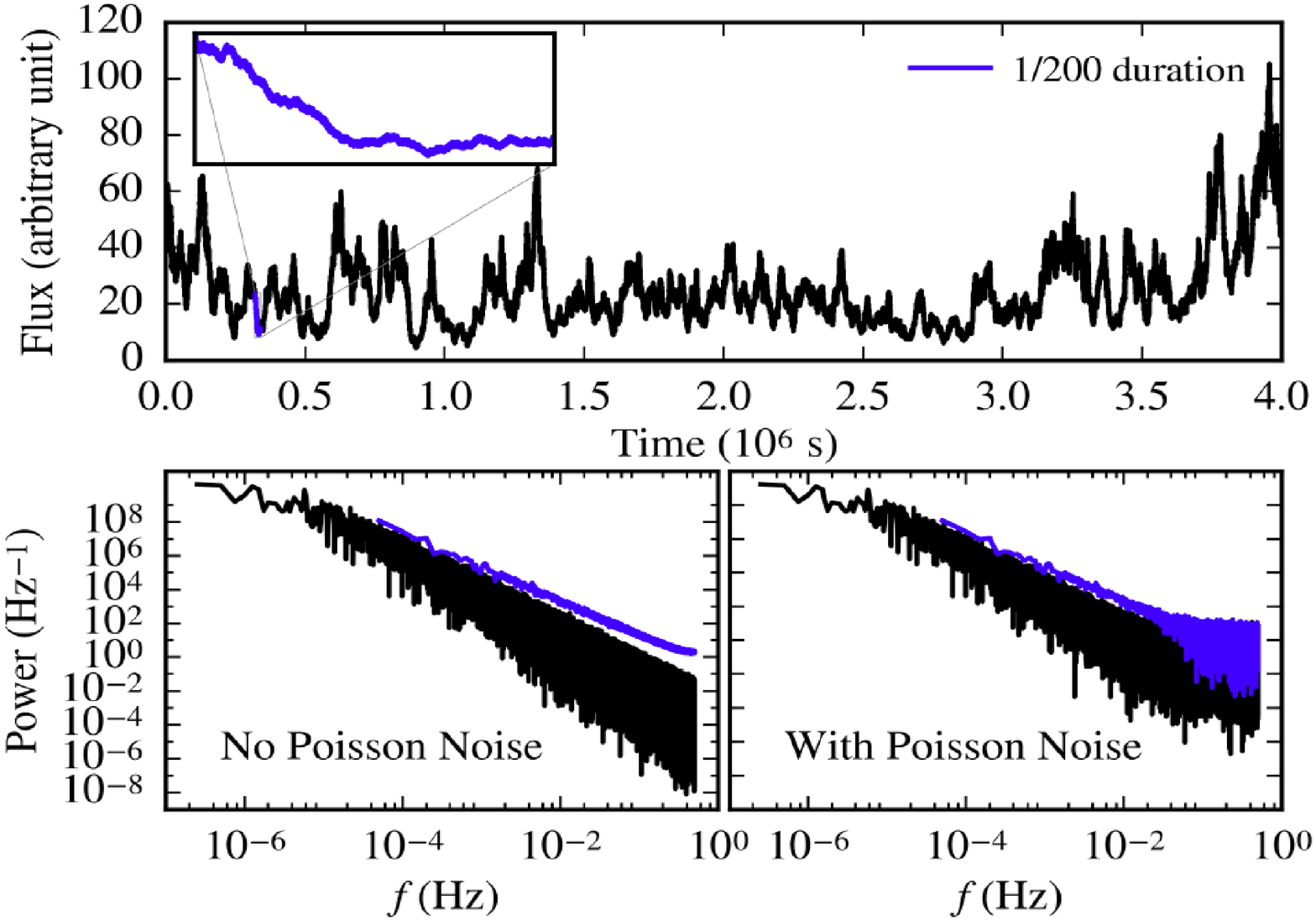}
 \caption{Top: Lightcurve simulated following the \cite{Uttley2005a} method, which possesses 
the properties of log-normal flux distribution and rms-flux relation that are found in
 real AGN lightcurves. The blue lightcurve in the inset is a segment of 1/200 length of the black lightcurve. 
Bottom left: Periodograms of the above two lightcurves without added Poisson noise.
Bottom right: Periodograms of the above two lightcurves with added Poisson noise, which behaves as white noise at the highest frequencies.
All periodograms are normalized with $A_{\textrm{rms}}$. Red-noise leakage is apparent in both bottom panels.}
\end{figure}

\begin{figure}
\figurenum{A2}
\label{fig:lognorm_poisson_NLS}
 \centering
 \includegraphics[width=0.48\textwidth, clip]{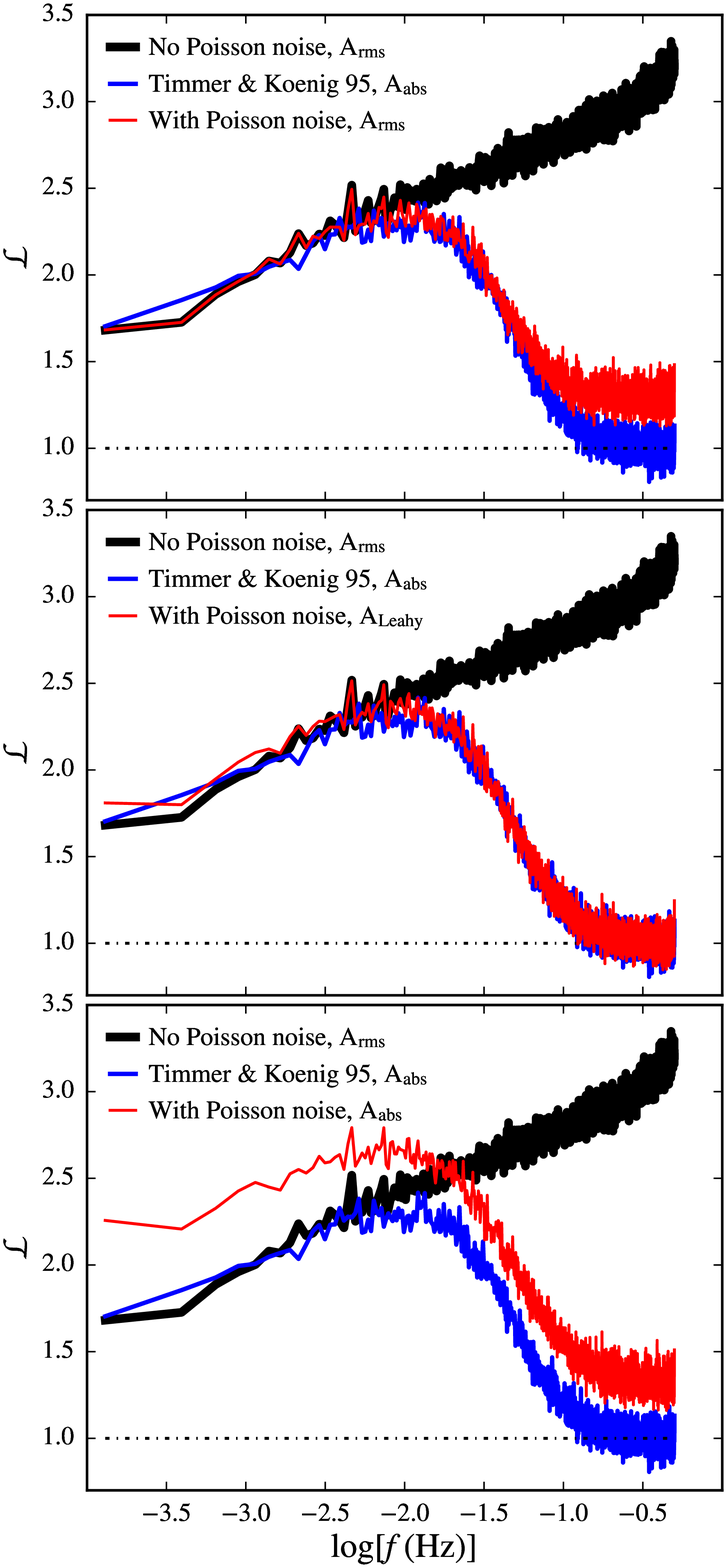}
 \caption{Comparison of NLS of different lightcurves with different normalizations adopted. 
 The NLS of the lightcurve with log-normal flux distribution but without
either Poisson or white noise is normalized
 with $A_{\textrm{rms}}$ and plotted in black in all panels. 
The NLS of the lightcurve with log-normal flux distribution and Poisson noise
is normalized in three
 different ways and plotted in red in different panels.
For comparison, the NLS produced from our previous simulation using the simpler \cite{Timmer1995} algorithm
is plotted in blue in all panels.} 
\end{figure}

When utilizing our NLS approach, as noted in Section~\ref{sec:discussion}, 
there are additional choices of periodogram normalization (i.e., $A_{\textrm{rms}}$ and $A_{\textrm{Leahy}}$,
other than $A_{\textrm{abs}}$) and some features of real AGN data to be further explored.
In this appendix, we therefore carry out an additional set of simulations
to investigate the effects of log-normal flux distribution and Poisson-statistics sampled lightcurves (i.e., features of real AGN data) on the performance of NLS.

Following the method of \cite{Uttley2005a} and using a slightly different PSD model from 
the one adopted in Section~\ref{sec:two_biases} and Fig.~\ref{fig:two_leakage},
we first generate an artificial lightcurve 
with the \cite{Timmer1995} algorithm that is briefly outlined in Section~\ref{sec:sim_lc}. 
We then take the exponential of the flux to produce the log-normal distribution and
exclude the high-frequency white-noise component of the PSD model in the simulation. 
Specifically, the PSD model used here is in the form of
\begin{equation}\label{eq:new_psd}
    \mathcal{P}(f) =
 \begin{cases}
 Af^{-\alpha} , & f>f_\textrm{c} \\
 Af_\textrm{c}^{-\alpha}\left({f}/{f_\textrm{c}}\right)^{-\beta}, & f<f_\textrm{c},
 \end{cases}
\end{equation}
where $\alpha=2.4$, $\beta=1.0$, and $f_{\textrm{c}}=10^{-5}$~Hz (i.e., compared to Eq.~\ref{eq:PSD_model},
the only difference is the missing high-frequency white-noise term).
Additionally, we sample the resultant lightcurve using
Poisson statistics to produce the errors in real observations.
The lightcurve is $4\times10^6$~s long and $\Delta T=1$~s.
The final lightcurve is plotted in the top panel of Fig.~\ref{fig:Uttley05_lc}, where the inset displays a $2\times10^4$~s segment.
Taking the exponential of a lightcurve with typical rms only changes slightly the shape of its broad continuum power spectrum \citep[see Appendix~B of][for a demonstration]{Uttley2005a}.
Despite of being generated with a nearly identical power spectral shape
(see Eq.~\ref{eq:new_psd} and Eq.~\ref{eq:PSD_model} for the slight difference between them),
the lightcurve here is quite different from the
lightcurve shown in Fig.~\ref{fig:two_leakage}, due to the usage of different simulation algorithms:
the former displays multiple flares that are largely above a non-flaring (``quiescent'') flux level, 
being very similar to real AGN lightcurves; 
while the latter displays both positive and negative fluctuations of similar amplitudes  
around the mean flux level.
The periodograms of the whole lightcurve ($4\times10^6$~s long) and the segment lightcurve
($2\times10^4$~s long), without and with added Poisson noise, 
are shown in the
bottom-left and bottom-right panels of Fig.~\ref{fig:Uttley05_lc}, respectively. 
These periodograms are normalized with $A_{\textrm{rms}}=2\Delta T/\bar{x}^2n$, where $\bar{x}$
represents the corresponding mean value of each of the above two lightcurves.
It is clear that, at the highest frequencies above $\sim10^{-2}$~Hz,
Poisson noise affects the periodograms in a similar way as white noise does 
(cf. the right panel of Fig.~\ref{fig:two_leakage}).

We then cut the $4\times10^6$~s lightcurve into 200 segments of equal length, 
calculate the NLS with different choices of normalization, and 
plot these NLS in Fig.~\ref{fig:lognorm_poisson_NLS}.
In each panel of Fig.~\ref{fig:lognorm_poisson_NLS},
we plot the same NLS of the lightcurve without added Poisson noise in black 
(i.e., no Poisson or white noise) and use only $A_{\textrm{rms}}$; 
for comparison, we also plot the same NLS in blue that is generated from our previous simulation with 
the exactly same configuration except for adding white noise
and is normalized by $A_{\textrm{abs}}=2\Delta T/n$ 
(see the NLS in panel (c) of Fig.~\ref{fig:red_leakage}).
Without added (Poisson) noise, the lightcurve does not contain any frequency-independent variation,
so its NLS (i.e., the black one) goes straight up without falling down,
due to that the bias curve goes straight up if there is no noise
(see Figs.~\ref{fig:red_leakage}a and \ref{fig:red_leakage}c);
nevertheless, this NLS slope is still consistent with the slope of the blue NLS
that is from the simpler simulation with the \cite{Timmer1995} algorithm.
In Fig.~\ref{fig:lognorm_poisson_NLS}, 
we also use three different normalizations to calculate the NLS of the lightcurve 
with added Poisson noise and plot them in red. 
In the top panel, we use
$A_{\textrm{rms}}$ for the red NLS, which has a slope consistent with that of the other two NLS
and falls to a level above 1.0 (rather than being 1.0 for the blue NLS) 
after Poisson noise starts to dominate the power spectrum at high frequencies.
In the middle panel, as we use $A_{\textrm{Leahy}}=2\Delta T/\bar{x}n$ to calculate the periodogram, the red NLS falls
and flattens to 1.0 eventually, but its slope is slightly
flatter than the other two NLS (note that a straightforward comparison between the slopes of
the red and black NLS can be made as they are from the same set of simulations).
In the bottom panel, 
we use $A_{\textrm{abs}}$ for the red NLS, which lies above the blue NLS at all frequencies and
above the black NLS before falling down, and has a flatter slope than the other two NLS (compared to 
the middle panel, the degree of the red NLS being flatter is more severe). 

It is not surprising to see the above dependence of NLS behavior on the different choices of normalization.
As we know, the absolute power of Poisson noise is proportional to $\bar{x}$ \citep{Vaughan2003} 
and the intrinsic variation amplitude is proportional to $\bar{x}^2$ 
due to the rms-flux relation \citep{Uttley2005a}, 
where $\bar{x}$ is the local mean of each segment lightcurve. 
Using $A_{\textrm{rms}}$ and $A_{\textrm{Leahy}}$ cancels out the
factors of $\bar{x}$ associated with the intrinsic variation amplitude and
the absolute power of Poisson noise, respectively; as the factors of $\bar{x}$ 
cannot be canceled out simultaneously, the red NLS is lifted up either at
high frequencies (see the top panel) or low frequencies (see the middle panel).
While using $A_\textrm{abs}$ cannot cancel out any dependence on the
mean flux level, the red NLS is lifted up at all frequencies compared to the blue NLS (see the bottom panel).

Based on the above analysis and the effectiveness of the method of using the NLS slope
to constrain the high-frequency PSD slope as demonstrated 
in Sections~\ref{sec:free1palpha} and \ref{subsec:fix_one_parameter},
we make the following recommendations regarding the usage of our NLS approach:
(1) it is fine to adopt $A_{\textrm{abs}}$ when dealing with lightcurves without realistic features
(i.e., the blue NLS);
(2) it is fine to adopt $A_{\textrm{rms}}$ or even $A_{\textrm{Leahy}}$ when dealing with
lightcurves with realistic features (i.e., the red NLS in the top and middle panels); and
(3) it is risky to adopt $A_{\textrm{abs}}$ when dealing with lightcurves with realistic features
(i.e., the red NLS in the bottom panel).
Finally, we conclude that, with an appropriate choice of normalization, the NLS can be used to 
effectively constrain the high-frequency PSD of lightcurves with or without realistic features, which 
demonstrates that the NLS is an effective and consistent tool to study AGN PSD.

\bibliographystyle{astron}
\bibliography{bibliography}

\end{document}